\documentclass[10pt,conference,a4paper]{IEEEtran}
\usepackage{times,amsmath,mathabx,epsfig}
\usepackage{cite}
\usepackage{booktabs}
\usepackage{multirow}

\newcommand{\be}{\begin{equation}}
\newcommand{\ee}{\end{equation}}
\newcommand{\eq}[1]{(\ref{#1})}
\newcommand{\Ss}{\mathcal{S}}

\title{SolarStat: Modeling Photovoltaic Sources \\ through Stochastic Markov Processes}
\author{%
{Marco Miozzo{$^\star$}, Davide Zordan{$^\dag$}, Paolo Dini{$^\star$} and Michele Rossi{$^\dag$} }%
\vspace{1.0mm}\\
\fontsize{10}{10}\selectfont\itshape
$^\star$ CTTC, Av. Carl Friedrich Gauss, 7, 08860, Castelldefels, Barcelona, Spain\\
\fontsize{9}{9}\selectfont\ttfamily\upshape
E-mail: {mmiozzo,pdini}@cttc.es\\
\vspace{1.0mm}\\
\fontsize{10}{10}\selectfont\rmfamily\itshape
$^\dag$\,DEI, University of Padova, Via G. Gradenigo, 6/B, 35131 Padova, Italy\\
\fontsize{9}{9}\selectfont\ttfamily\upshape
E-mail: {zordanda, rossi}@dei.unipd.it}
\begin{document}
\maketitle
\begin{abstract}
In this paper, we present a methodology and a tool to derive simple but yet accurate stochastic Markov processes for the description of the energy scavenged by outdoor solar sources. In particular, we target photovoltaic panels with small form factors, as those exploited by embedded communication devices such as wireless sensor nodes or, concerning modern cellular system technology, by small-cells. Our models are especially useful for the theoretical investigation and the simulation of energetically self-sufficient communication systems including these devices. 

The Markov models that we derive in this paper are obtained from extensive solar radiation databases, that are widely available online. Basically, from hourly radiance patterns, we derive the corresponding amount of energy (current and voltage) that is accumulated over time, and we finally use it to represent the scavenged energy in terms of its relevant statistics. Toward this end, two clustering approaches for the raw radiance data are described and the resulting Markov models are compared against the empirical distributions. 

Our results indicate that Markov models with just two states provide a rough characterization of the real data traces. While these could be sufficiently accurate for certain applications, slightly increasing the number of states to, e.g., eight, allows the representation of the real energy inflow process with an excellent level of accuracy in terms of first and second order statistics.  

Our tool has been developed using Matlab\texttrademark  and is available under the GPL license at~\cite{Solar_stat_code}.
\end{abstract}


\section{Introduction}
The use of renewable energy is very much desirable at every level of the society, from industrial / manufacturing activities to smart cities, public buildings, etc. Being able to capture any sort of renewable energy source is in fact very useful to power up, e.g., sensing equipment and electric apparatuses that surround us in our daily life, from automatic doors, to sensor systems for traffic control, intrusion detection, alarms, pollution reporting, etc. According to the paradigm of ÒSmart CitiesÓ and the Internet of Things~\cite{Morabito-IoT-10,SmartCities-2013}, these ``resource constrained'' small sensing devices are expected to be deployed massively. Of course, as an immediate advantage, self-sufficient (also referred to as {\it perpetual}) networks that will live unattended, just thanks to the energy they scavenge from the environment, would cut down their maintenance cost. Moreover, these systems will contribute to the reduction of the energy absorbed from the power grid (which is usually obtained 
from carbon fossil or nuclear power plants), thus benefiting the environment. 

Similarly, Cellular Networks are expecting a tremendous grow of the traffic demand in the next years with a consequent increase in terms of cost and energy consumption. A challenging but promising solution is represented by the deployment of Base Stations (BSs) employing renewable energy sources~\cite{Chen-2011}. 
Note, however, that the mere integration of a solar panel into existing electrical apparatuses, such as macro BSs, is often not sufficient as keeping these devices fully operational at all times would demand for unrealistically large solar modules~\cite{Piro-2013}. To overcome this, the energy coming from the renewable sources should be wisely used, predicting future energy arrival and the energy consumption that is needed by the system to remain operational when needed. This calls for complex optimization approaches that will adapt the behavior of modern systems to the current application needs as well as to their energy reserves and the (estimated) future energy inflow~\cite{Gunduz-2013}.

A large body of work has been published so far to mathematically analyze these facts, especially in the field of wireless sensor networks. However, often researchers have tested their ideas considering deterministic~\cite{Ozel-2011,Gregori-TW-2013}, iid distributed across time slots~\cite{Gatzianas-2010} or time-correlated Markov models~\cite{Michelusi-TCOM-2013}. While these contributions are valuable for the establishment of the theory of energetically self-sufficient  networks; seldom, the actual energy production process in these papers has been linked to that of real solar sources, to estimate the effectiveness of the proposed strategies under realistic scenarios.   

The work in this paper aims at filling this gap, by providing a methodology and a tool to obtain simple but yet accurate stochastic Markov processes for the description of the energy scavenged by outdoor solar sources. In this study, we focus on solar modules as those that are installed in wireless sensor networks or small-LTE cells, by devising suitable Markov processes with first- and second-order statistics that closely match that of real data traces. Our Markov models allow the statistical characterization of solar sources in simulation and theoretical developments, leading to a higher degree of realism.  

This paper is organized as follows. In Section~\ref{sec:system_model} we detail the system model and in particular how the raw radiance data is processed to estimate the corresponding instantaneous harvested power. This requires the combination of several building blocks, including  an astronomical model (Section~\ref{sec:sun}) to estimate the actual irradiance that hits the solar module, given the inclination of the sun during the day and the module placement, an electrical model of photovoltaic cells (Section~\ref{sec:PV_module}) and a model for the DC/DC power processor (Section~\ref{sec:power_processor}), which is utilized to maximize the amount of power that is collected. Hence, in Section~\ref{sec:energy_model} we describe the Markov model that we use to statistically describe the energy inflow, according to two clustering approaches for the raw data. The results from this Markov model are shown in Section~\ref{sec:results}, whereas our conclusions are presented in Section~\ref{sec:conclusions}.  

\section{System Model}
\label{sec:system_model}

\begin{figure*}
\begin{center}
\includegraphics[width=0.65\textwidth]{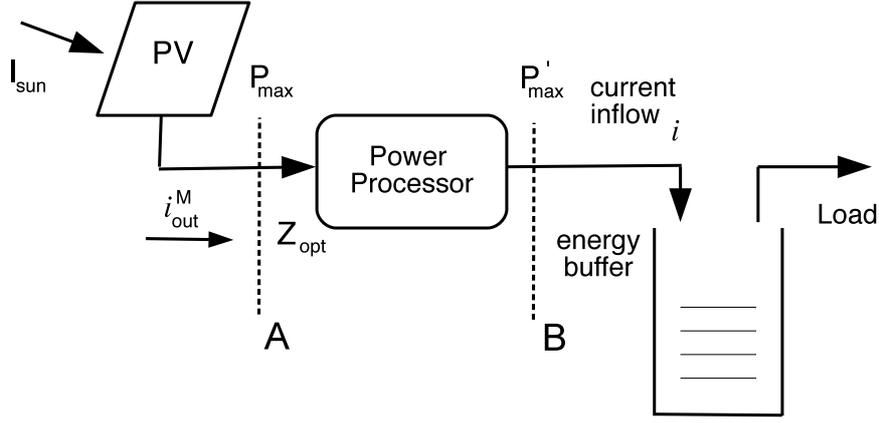}
\caption{Diagram of a solar powered device. The power processor adapts its input impedance so that it will match that of the source, $Z_{\rm opt}$. This allows the extraction of the maximum power $P_{\max}$.} 
\label{fig:BS_model}
\end{center}
\end{figure*}

In this section, we describe the source model that we have adopted to statistically describe the energy inflow for a solar powered embedded device, see also~\cite{Culler-Harvesting-12}. To facilitate our description, we consider the diagram of Fig.~\ref{fig:BS_model} where we identify the key building blocks for our study: the solar source (indicated as $I_{\rm sun}$), the photovoltaic panel (PV), the DC/DC power processor and the energy buffer (i.e., a rechargeable battery). In Section~\ref{sec:sun} we start with the characterization of the effective solar irradiance, $I_{\rm eff}$, that in general depends on the geographical coordinates of the installation site, the season of the year and the hour of the day. Hence, $I_{\rm eff}$ is translated by the PV module into some electrical power and a DC/DC power processor is used to ensure that the maximum power is extracted from it. 

\subsection{Astronomical Model}
\label{sec:sun}

The effective solar radiance that hits a photovoltaic module, $I_{\rm eff}$, depends on physical factors such as its location, the inclination of the solar module, the time of the year and the hour of the day. Solar radiation databases are available for nearly all locations around the Earth and their data can be used to obtain the statistics of interest. An astronomical model is typically utilized to translate the instantaneous solar radiance $I_{\rm sun}$ (expressed in W/m$^2$) into the effective sunlight that shines on the solar module. According to~\cite{Dave-75}, the effective solar radiance that hits the solar module, $I_{\rm eff}$, is proportional to $\cos \Theta$, where $\Theta \in [-90^\degree,90^\degree]$ is the angle between the sunlight and the normal to the solar module surface\footnote{$\Theta=0$ ($\Theta=\pm 90^\degree$) if the sunlight arrives perpendicular (parallel) to the module.}. Astronomical models can be found in, e.g.,~\cite{Dave-75} and Chapter 8 of~\cite{Handbook-Energy-Technology-
2011}.

In short, $I_{\rm eff}$ depends on many factors such as the elliptic orbit of the Earth around the sun (which causes a variation of the distance between Earth and sun across different seasons), the fact that the Earth is itself tilted on its axis at an angle of $23.45^\degree$. This gives rise to a {\it declination} angle $\gamma$, which is the angular distance North or South of the Earth's equator, which is obtained as:
\be
\gamma(N) \simeq \sin^{-1} \left [ \sin(23.45^\degree) \sin \left ( D(N) \right ) \right ] \, , 
\ee
where $D(N) = 360 (N-81)/365^\degree$ and $N$ is the day number in a year with first of January being day $1$. Other key parameters are the {\it latitude} $La \in [0,90^\degree]$ (positive in either hemisphere), the {\it longitude} $Lo$, the {\it hour angle} $\omega(t,N) \in [0,360^\degree]$, that corresponds to the azimuth's angle of the sun's rays due to the Earth's rotation, the inclination $\beta$ of the solar panel toward the sun on the horizon and the azimuthal displacement $\alpha$, which is different from zero if the normal to the plane of the solar module is not aligned with the plane of the corresponding meridian, that is, the solar panel faces West or East.\footnote{$\alpha>0$ if the panel faces West and $\alpha<0$ if it faces East.} $\omega(t,N)$ is given by $\omega(t,N)=15(AST(t,N)-12)^\degree$, where $AST(t,N) \in [0,24]$ hours, is the apparent solar time, which is the time based on the rotation of the Earth with respect to the sun and is obtained as a scaled version of the local standard time 
$t$ (we refer to $t^\prime$ as $t$ adjusted accounting for the daylight savings time) for the time zone where the solar module is installed. $AST(t,N)$ is computed as follows. Briefly, we obtain the Greenwich meridian angle, $GMA = UTC_{\rm off} \times 15^\degree$, which corresponds to the angle between the Greenwich meridian and the meridian of the selected time zone: $UTC_{\rm off}$ is the time offset between Greenwich and the time zone and $15$ is the rotation angle of the Earth per hour. Thus, we compute $\Delta t = (Lo - GMA)/15^\degree$, i.e., the time displacement between the selected time zone and the time at the reference Greenwich meridian. At this point, $AST(t,N)$ is obtained as $AST(t,N) = t^\prime + \Delta t + ET(N)$ (expressed in hours), where $ET(N)$ is known as the {\it equation of time}, with $ET(N) \simeq [9.87 \sin (2D(N)) - 7.53 \cos(D(N)) -1.5 \sin(D(N))]/60$. 

Finally, the power incident on the PV module depends on the angle $\Theta$, for which we have:
\begin{eqnarray}
\cos \Theta(t,N) \!\!\! & = & \!\!\!  \sin \gamma(N) \sin La \cos \beta - \nonumber \\
& - & \sin \gamma(N) \cos La \sin \beta \cos \alpha + \nonumber \\ 
& + & \!\!\!  \cos \gamma(N) \cos La \cos \beta \cos \omega(t,N) +  \nonumber \\
& + & \!\!\!  \cos \gamma(N) \sin La \sin \beta \cos \alpha \cos \omega(t,N) \nonumber \\
& + & \!\!\!  \cos \gamma(N) \sin \beta \sin \alpha \sin \omega(t,N) \, .
\end{eqnarray}
Once an astronomical model is used to track $\Theta$, the effective solar radiance as a function of time $t$ is given by: $I_{\rm eff}(t,N) = I_{\rm sun}(t,N) \max(0, \cos \Theta(t,N))$, where the $\max(\cdot)$ accounts for the cases where the solar radiation is above or below the horizon, as in these cases the sunlight arrives from below the solar module and is therefore blocked by the Earth. The sun radiance, $I_{\rm sun}(t,N)$, for a given location, time $t$ and day $N$, has been obtained from the database at~\cite{nrel}.

\subsection{PV Module}
\label{sec:PV_module}

A PV module is composed of a number $n_{\rm sc}$ of {\it solar cells} that are electrically connected according to a certain configuration, whereby a number $n_{\rm p}$ of them are connected in parallel and $n_{\rm s}$ in series, with $n_{\rm sc} = n_{\rm p} n_{\rm s}$. A given PV module is characterized by its I-V curve, which emerges from the composition of the I-V curves of the constituting cells. Specifically, the I-V curve of the single solar cell is given by the superposition of the current generated by the solar cell diode in the dark with the so called {\it light-generated} current $i_\ell$~\cite{Lindholm-79}, where the latter is the photo-generated current, due to the sunlight hitting the cell. The I-V curve of a solar cell can be approximated as:
\be
\label{eq:IVcurve}
i_{\rm out} \simeq i_{\ell}  - i_{\rm o} \left [ \exp \left ( \frac{q v}{n k T}Ê\right ) - 1 \right ] \, ,
\ee 
where $q$ is the elementary charge, $v$ is the cell voltage, $k$ is the Boltzmann's constant, $T$ is the temperature in degree Kelvin\footnote{$T$ is given by the sum of the ambient temperature, which can be obtained from the dew point and relative humidity, and of a further factor due to the solar power hitting the panel.}, $n \geq 1$ is the diode ideality factor and $i_{\rm o}$ is the {\it dark saturation current}. $i_o$ corresponds to the solar cell diode leakage current in the absence of light and depends on the area of the cell as well as on the photovoltaic technology. The open circuit voltage $v_{\rm oc}$ and the short circuit current $i_{\rm sc}$ are two fundamental parameters for a solar cell. The former is the maximum voltage for the cell and occurs when the net current through the device is zero. $i_{\rm sc}$ is instead the maximum current and occurs when the voltage across the cell is zero (i.e., when the solar cell is short circuited). If $v_{\rm oc}^{\rm M}$ and $i_{\rm sc}^{\rm M}$ are the 
open circuit voltage and short circuit current for a solar module M, the single solar cell parameters are obtained as: $i_{\rm sc} = i_{\rm sc}^{\rm M} / n_{\rm p}$ and $v_{\rm oc} = v_{\rm oc}^{\rm M} / n_{\rm s}$ (considering a module composed of homogeneous cells).

The light-generated current for the single solar cell is a time varying quantity, $i_{\ell}(t,N)$, which depends on the amount of sunlight that hits the solar cell at time $t$, where $N$ is the day number. Here, we have used the following relation: $i_{\ell}(t,N) = i_{\rm sc} F(t,N)$, where the {\it radiation rate} $F(t,N) \in [0,1]$ is obtained as $F(t,N) = 0.001 I_{\rm eff}(t,N)$, i.e., normalizing the effective irradiance hitting the solar cell with respect to the maximum radiation of $1$~kW/m$^2$ (referred to in the literature as ``one sun''~\cite{PV-Book-03}). Hence, $i_{\ell}(t,N)$ is plugged into \eq{eq:IVcurve} to obtain $i_{\rm out}(t,N)$ for a single solar cell as a function of the time $t$ for day $N$. The total current that is extracted from the solar module is: $i_{\rm out}^{\rm M}(t,N) = n_{\rm p} i_{\rm out}(t,N)$.

\subsection{Power Processor}
\label{sec:power_processor}

Generally speaking, every voltage or current source has a {\it maximum power point}, at which the average power delivered to its load is maximized. For example, a Th\'{e}venin voltage source delivers its maximum power when operating on a resistive load whose value matches that of its internal impedance. However, in general the load of a generic device does not match the optimal one, which is required to extract the maximum power from the connected solar source. To cope with this, in practice the optimal load is emulated through a suitable {\it power processor}, whose function is that of ``adjusting'' the source voltage (section A of Fig.~\ref{fig:BS_model}) until the power extracted from it is maximized,\footnote{This corresponds to adapting the input impedance of the power processor to $Z_{\rm opt}=Z_{\rm source}^*$, where $^*$ indicates the complex conjiugate.} which is also known as maximum power point tracking (MPPT). Ideally, through MPPT, the maximum output power is extracted from the solar panel under 
any given temperature and irradiance condition, adapting to changes in the light intensity. Commercially available power processors use ``hill climbing techniques''; as an example, in~\cite{Mattavelli-10} the authors propose advanced control schemes based on the downhill simplex algorithm, whereby the voltage and the switching frequency are jointly adapted for fast convergence to the maximum power point. See also~\cite{Ropp-03} for further information on MTTP algorithms and their comparative evaluation and~\cite{Brunelli-2008} for a low-power design targeted to wireless sensor nodes. In the present work, we have taken into account the DC/DC power processor by computing the operating point $(i_{\rm out}^{\rm M}, v^{\rm M})$ (see~\eq{eq:IVcurve}) for which the extracted power in section A, $P = i_{\rm out}^{\rm M} v^{\rm M}$, is maximized. Note that, if $i_{\rm out}$ and $v$ are the output current and the voltage of the single solar cell, we have $i_{\rm out}^{\rm M} = n_{\rm p} i_{\rm out}$ and $v^{\rm M} 
= n_{\rm s} v$. For this procedure, we have considered the parameters of Section~\ref{sec:sun} and~\ref{sec:PV_module} (solar irradiance, rotation of the Earth, etc.) and also the fact that $i_{\rm sc}$ and $v_{\rm oc}$ change as a function of the environmental temperature, which affects the shape of the I-V curve~\eq{eq:IVcurve} (see, e.g., the dependence of $i_{\ell}$ on $i_{\rm sc}$). Hence, we have computed the extracted power in two steps: step 1) we have obtained the (ideal) maximum power $P_{\rm MPP}$ that would be extracted by the panel at the MPP by an ideal tracking system:
\be
P_{\rm MPP} = \max_{v} \{ i_{\rm out}^{\rm M} v^{\rm M} \} = n_{\rm p} n_{\rm s} \max_{v} \{ i_{\rm out} v \} \, ,
\ee 
where $i_{\rm out}$ is given by \eq{eq:IVcurve}. Step 2) the power available after the power processor (section B in Fig.~\ref{fig:BS_model}) is estimated as $P_{\max}^\prime = \eta P_{\rm MPP}$, where $\eta \in (0,1)$ is the power processor conversion efficiency, which is usually defined as the ratio $P_{\max}^\prime / P_{\rm MPP}$ and can be experimentally characterized for a given MPP tracking circuitry~\cite{Brunelli-2008}. $P_{\max}^\prime$ is the power that is finally transferred to the energy buffer.

\subsection{Semi-Markov Model for Stochastic Energy Harvesting}
\label{sec:energy_model}

The dynamics of the energy harvested from the environment is captured by a continuous time Markov chain with $N_{\rm s} \geq 2$ states. This model is general enough to accommodate different clustering approaches for the empirical data, as we detail shortly.

Formally, we consider an energy source that, at any given time, can be in any of the states $x_s \in \Ss = \{0,1,\dots,N_{\rm s}-1\}$. We refer to $t_k$, with $k \geq 0$, as the time instants where the source transitions between states, and we define $\tau_k = t_{k+1} - t_k$ as the time elapsed between two subsequent transitions. In what follows, we say that the system between $t_k$ and $t_{k+1}$ is in {\it cycle} $k$.

Right after the $k$-th transition to state $x_s(k)$, occurring at time $t_k$, the source remains in this  state for $\tau_k$ seconds, where $\tau_k$ is governed by the probability density function (pdf) $f(\tau|x_s)$, with $\tau \in [\tau_{\min}(x_s),\tau_{\max}(x_s)]$. At the next transition instant, $t_{k+1}$, the source moves to state $x_s(k+1) \in \Ss$ according to the probabilities $p_{u v}=\textrm{Prob}\{x_s(k+1)=v | x_s(k)=u\}$, with $u,v \in \Ss$. 
When the source is in state $x_s(k)$, an input current $i_k$ is fed to the rechargeable battery, where $i_k$ is drawn from the pdf $g(i |x_s)$, with $i \geq 0$. That is, when a state is entered, the input current $i$ and the permanence time $\tau$ are respectively drawn from $g(i | x_s)$ and $f(\tau|x_s)$. Then, the input current remains constant until the next transition, that occurs after $\tau$ seconds. In this work, we assume that the voltage at the energy buffer (section B of Fig.~\ref{fig:BS_model}) is constant, as typically considered when a rechargeable battery is used. Given that, there is a one-to-one mapping between instantaneous harvested power and harvested current.

\subsection{Estimation of Energy Harvesting Statistics}
\label{sec:statistics}

Based on our models of Sections~\ref{sec:sun}-\ref{sec:power_processor}, we have mapped the hourly irradiance patterns obtained from~\cite{nrel} into the corresponding operating point, in terms of power $P_{\max}^\prime$ and current $i$ after the power processor (section B of Fig.~\ref{fig:BS_model}). Thus, we have computed the statistics $f(\tau | x_s)$ and $g(i | x_s)$ of Section~\ref{sec:energy_model} from these data according to the two approaches that we describe next. These differ in the adopted clustering algorithm, in the number of states $N_{\rm s}$ and in the structure of the transition probabilities $p_{uv}$, $u,v \in \Ss$.\\

\noindent \textbf{Night-day clustering:} we have collected all the data points in~\cite{nrel} from 1991 to 2010 and  grouping them by month. Thus, for each day in a given month we have classified the corresponding points into two states $x_s \in \{0,1\}$, i.e., a low- ($x_s=1$) and a high-energy state ($x_s=0$). To do this, we have used a current threshold $i_{\rm th}$, which is a parameter set by the user, corresponding to a small fraction of the maximum current in the dataset. According to the resulting value of $i_{\rm th}$, we have classified all the points that fall below that threshold as belonging to state $0$ (i.e., night) and those points above the threshold as belonging to state $1$ (day). After doing this for all the days in the dataset, we have estimated the probability density function (pdf) of the duration $\tau$, $f(\tau | x_s)$, and that of the input current $i$ (after the power processor), $g(i | x_s)$, for each state and for all  months of the year. For the estimation of the pdfs we have used the kernel smoothing technique see, e.g.,~\cite{Smoothing-book-96}. The transition probabilities of the resulting semi-Markov chain are $p_{10}=p_{01}=1$ and $p_{00}=p_{11}=0$ as a night is always followed by a day and vice versa.\\

\noindent \textbf{Slot-based clustering:} as above, we have collected and classified the irradiance data by month. Then, we subdivided the $24$ hours in each day into a number $N_{\rm s} \geq 2$ of time slots of constant duration, equal to $T_i$~hours, $i=1,\dots,N_{\rm s}$. Each slot is a state $x_s$ of our Markov model. Hence, for each state $x_s$ we computed the pdf $g(i | x_s)$ for each month of the year, considering the empirical data that has been measured in slot $x_s$ for all days in the dataset for the month under consideration. Again, the kernel smoothing technique has been utilized to estimate the pdf. For the statistics $f(\tau|x_s)$, being the slot duration constant by construction, we have that: $f(\tau|x_s) = \delta(\tau-T_{x_s})$, for all states $x_s \in \Ss$, where $\delta(\cdot)$ is the Dirac's delta.  The transition probabilities of the resulting Markov chain are $p_{u v} = 1$, when $u \in \Ss$ and $v = (u+1) \mod N_{\rm s}$, and $p_{u v} = 0$ otherwise. This reflects the temporal arrangement of the states.


\section{Numerical Results}
\label{sec:results}

For the results in this section, we have used as reference the commercially available micro-solar panels from Solarbotics, selecting the Solarbotics's SCC-3733 Monocrystalline solar technology~\cite{Solarbotics}. For this product, the single cell area is about $1$~square centimeters, the solar cells have an efficiency of $21.1$\%, $i_{\rm sc}=5$~mA and $v_{\rm oc}=1.8$~V. For the DC/DC power processor we have set $\eta=0.5$ and $V_{\rm ref}=3$~V, which are typical values for embedded sensor nodes, see~\cite{Brunelli-2008} and~\cite{Z1-Platform}. Next, we show some results on the stochastic model for the solar energy source of Section~\ref{sec:system_model}. These are obtained considering a solar module with $n_{\rm p}=6$ and $n_{\rm s}=6$ cells in parallel and in series, respectively. We have selected Los Angeles as the installation site, considering $\beta=45^\degree$, $\alpha=30^\degree$ and processing the data from~\cite{nrel} as described in the previous section with a cluster threshold equal to $1/50{\rm -th}$ of the maximum value of the current in the dataset.\\

\begin{figure}[t]
\begin{center}
\includegraphics[width=\columnwidth]{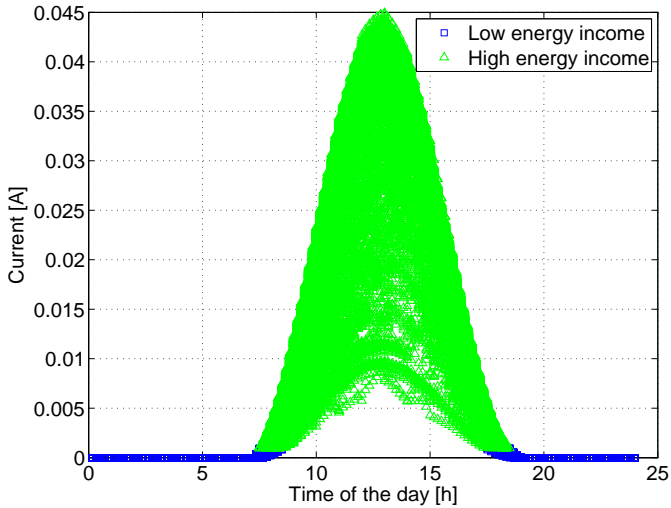}
\caption{Result of the night-day clustering approach for the month of July considering the radiance data from years $1999-2010$.} 
\label{fig:nighf-day-clustering-july}
\end{center}
\end{figure}

\begin{figure}[t]
\begin{center}
\includegraphics[width=\columnwidth]{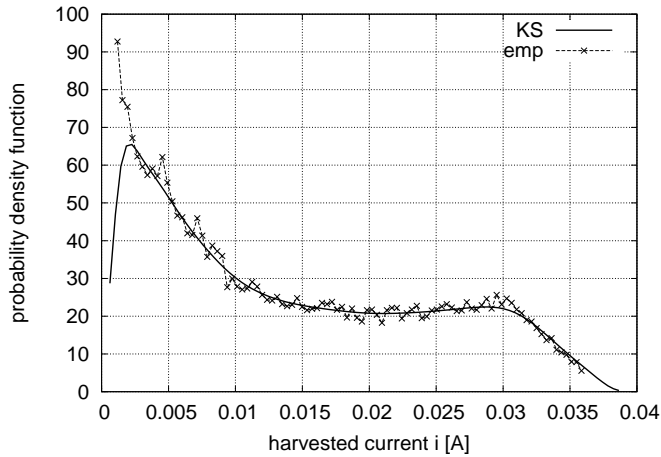}
\caption{$g(i | x_s)$ (solid line, $x_s=0$) obtained through the Kernel Smoothing (KS) technique for the month of February, for the night-day clustering method (2-state semi-Markov model), using radiance data from years $1999-2010$. The empirical pdf (emp) is also shown for comparison.} 
\label{fig:pdf_emp}
\end{center}
\end{figure}

\noindent {\bf Night-day clustering:} a first example for the night-day clustering approach is provided in Fig.~\ref{fig:nighf-day-clustering-july}, which shows the result of the clustering process for the month of July. Two macro states are evident: a low energy state (night), during which the power inflow is close to zero, and a high energy state (day). As this figure shows, the harvested current during the day follows a bell-shaped curve. However, contrarily to what one would expect, even for the month of July the high-energy state shows a high degree of variability from day-to-day, as is testified by the considerable dispersion of points across the y-axis. This reflects the variation in the harvested current due to diverse weather conditions. In general we have a twofold effect: (i) for different months the peak of the bell varies substantially, e.g., from winter to summer and (ii) for a given month the variability across the y-axis remains among different days. These facts justify the use of 
stochastic modeling, as 
we do in this work, to capture such variability in a statistical sense.  

Another example, regarding the accuracy of the Kernel Smoothing (KS) technique to fit the empirical pdfs, is provided in Fig.~\ref{fig:pdf_emp}, where we show the fitting result for the month of February.

\begin{figure}
 \begin{minipage}[b]{8.5cm}
   \centering
   \includegraphics[width=\columnwidth]{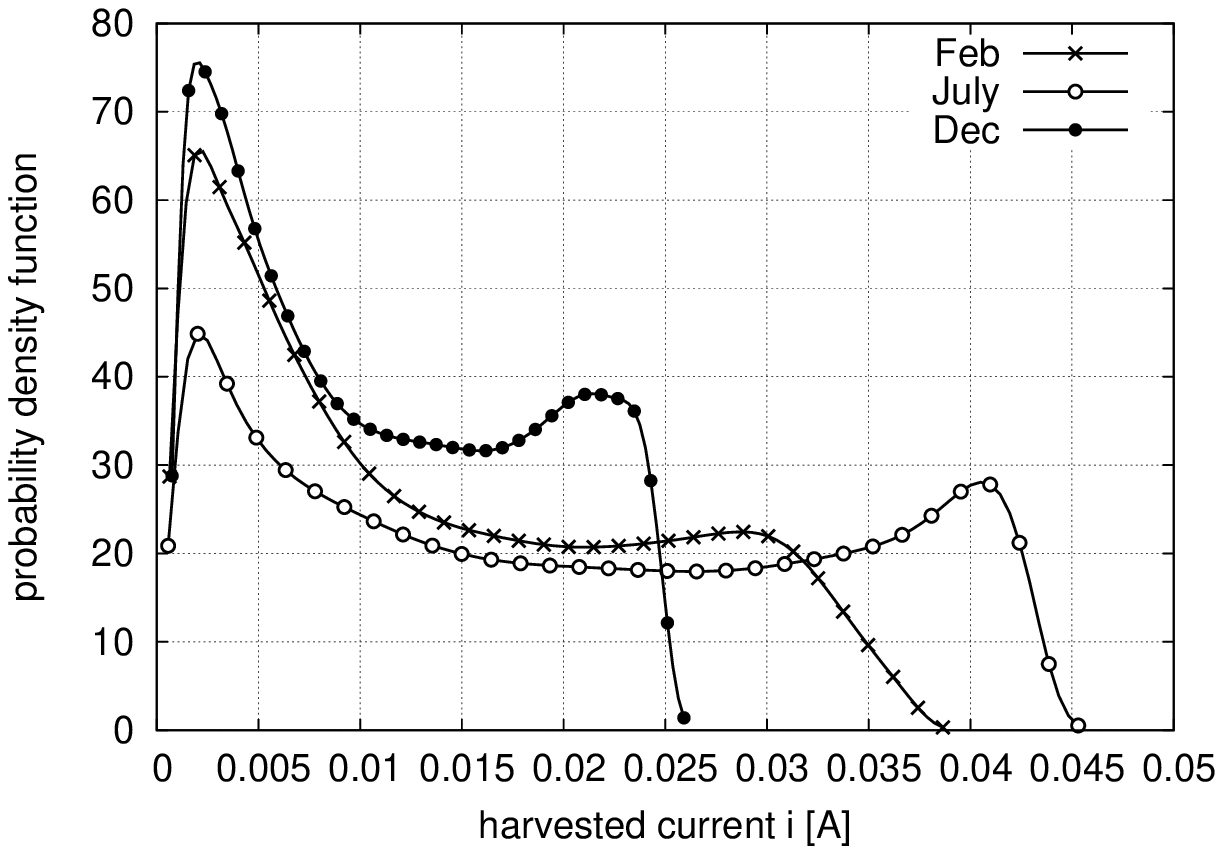}
   \caption{Probability density function $g(i | x_s)$, for $x_s = 1$, obtained through Kernel Smoothing for the night-day clustering method (2-state Markov model).}
   \label{fig:pdf_1}
 \end{minipage}
 \ \hspace{2mm} \hspace{3mm} \
 \begin{minipage}[b]{8.5cm}
  \centering
  \includegraphics[width=\columnwidth]{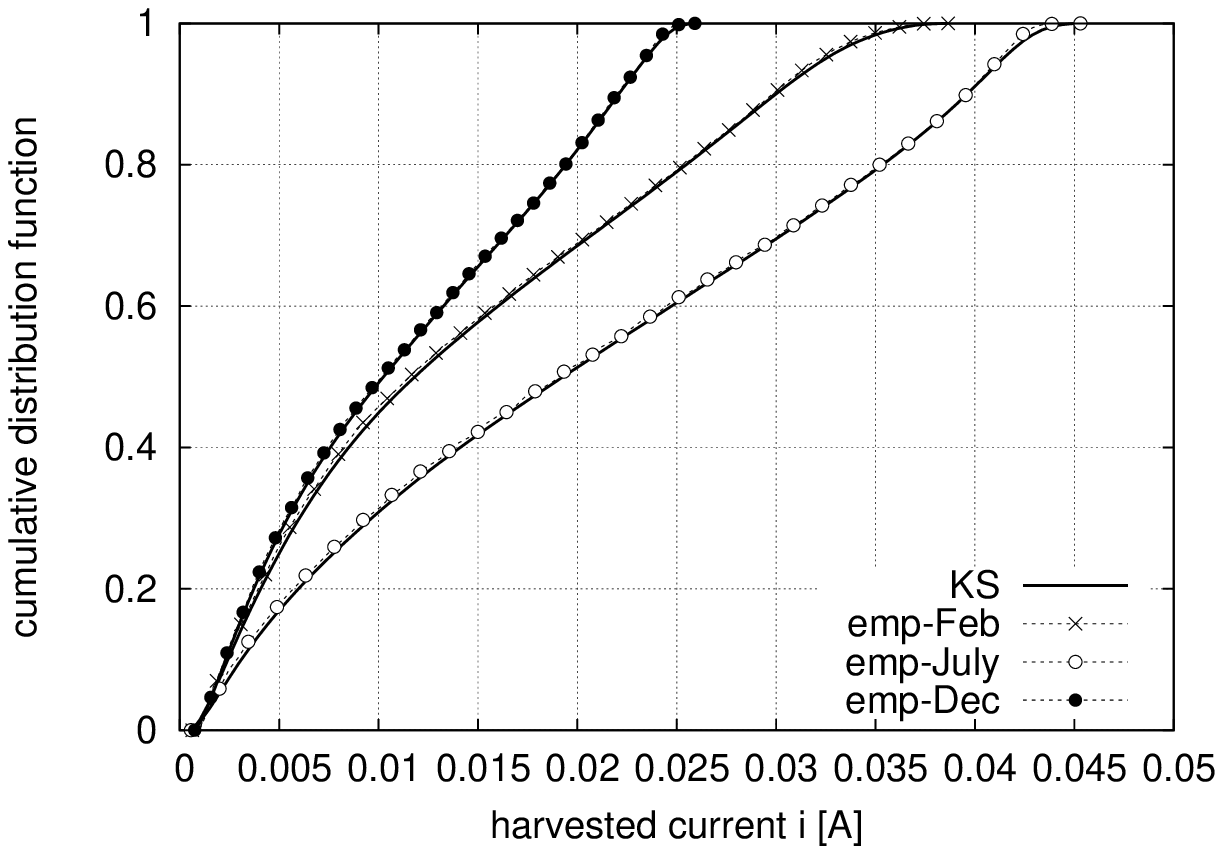}
   \caption{Cumulative distribution function of the harvested current for $x_s = 1$ (solid lines), obtained through Kernel Smoothing (KS) for the night-day clustering method (2-state Markov model). Empirical cdfs (emp) are also shown for comparison.}
   \label{fig:cdf_1}
 \end{minipage}
\end{figure}

\begin{figure}
 \begin{minipage}[b]{8.5cm}
   \centering
   \includegraphics[width=\columnwidth]{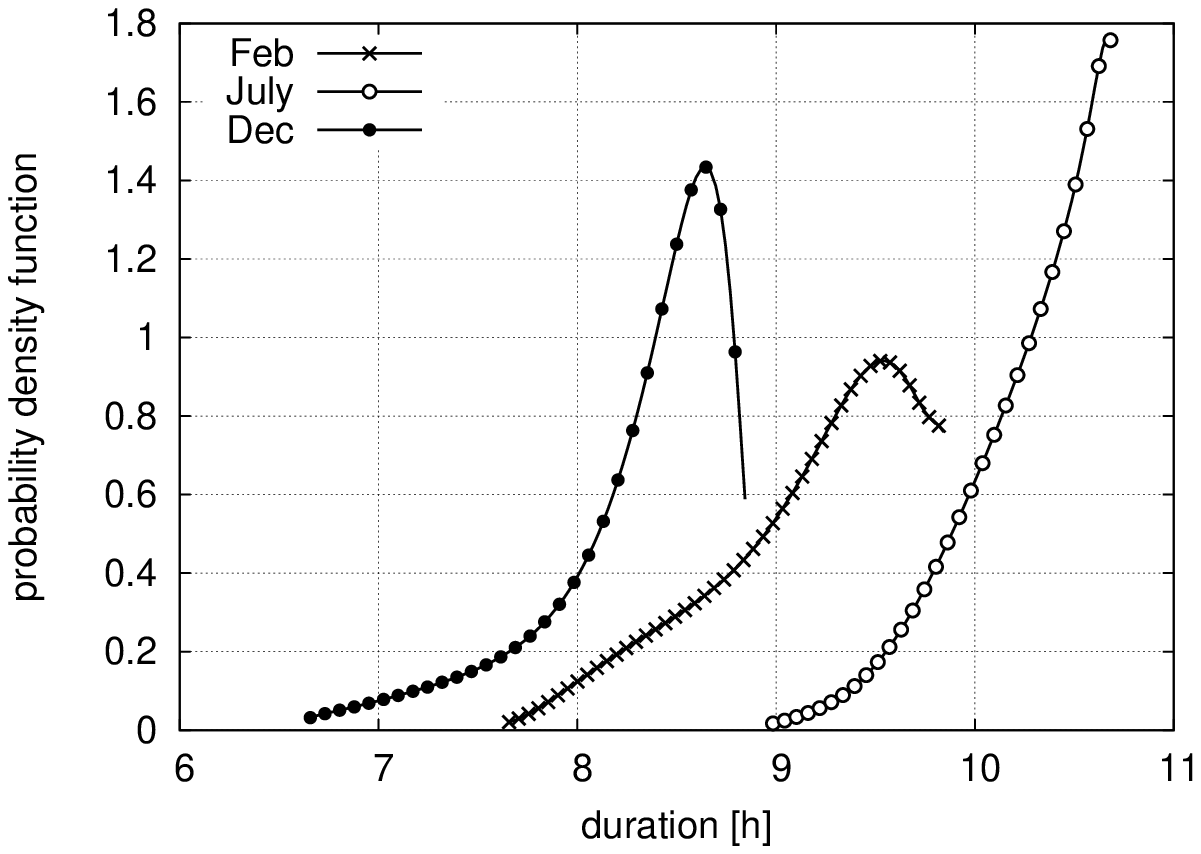}
   \caption{Probability density function $f(\tau | x_s)$, for $x_s = 1$, obtained through Kernel Smoothing for the night-day clustering method (2-state Markov model).}
   \label{fig:pdf_2}
 \end{minipage}
 \ \hspace{2mm} \hspace{3mm} \
 \begin{minipage}[b]{8.5cm}
  \centering
  \includegraphics[width=\columnwidth]{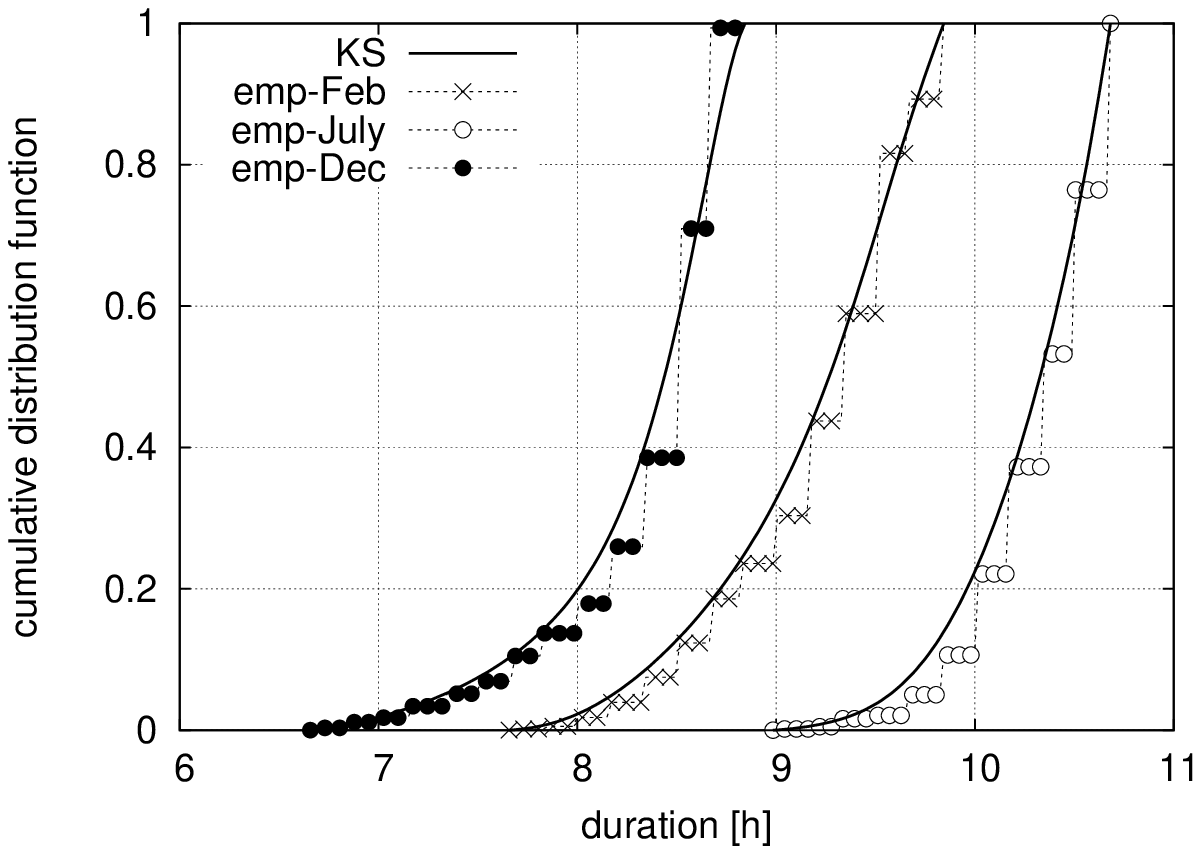}
   \caption{Cumulative distribution function of the state duration for $x_s = 1$ (solid lines), obtained through Kernel Smoothing (KS) for the night-day clustering method (2-state Markov model). Empirical cdfs (emp) are also shown for comparison.}
   \label{fig:cdf_2}
 \end{minipage}
\end{figure}

In Figs.~\ref{fig:pdf_1} and \ref{fig:cdf_1} we show some example statistics for the months of February, July and December. In Fig.~\ref{fig:pdf_1}, we plot the pdf $g(i|x_s)$, which has been obtained through the Kernel Smoothing (KS) technique for the high-energy state $x_s=0$. As expected, the pdf for the month of July has a larger support and has a peak around $i=0.04$~A, which means that is likely to get a high amount of input current during that month. For the months of February and December, we note that their supports shrink and the peaks move to the left to about $0.03$~A and $0.022$~A, respectively, meaning that during these months the energy scavenged is lower and is it more likely to get a small amount of harvested current during the day. Fig.~\ref{fig:cdf_1} shows the cumulative distribution functions (cdf) obtained integrating $g(i|x_s)$ and also the corresponding empirical cdfs. From this graph we see that the cdfs obtained through KS closely match the empirical ones. In particular, all the 
cdfs that we have obtained through KS have passed the Kolmogorov-Smirnov test when compared against the empirical ones, for a confidence of $1$\%, which confirms that the obtained distributions represent a good model for the statistical characterization of the empirical data. The pdf for state $x_s=1$ is not shown as it has a very simple shape, presenting a unique peak around $i = 0^+$. In fact, the harvested current is almost always negligible during the night.\footnote{Note that our model does not account for the presence of external light sources such as light poles.}  
Figs.~\ref{fig:pdf_2} and \ref{fig:cdf_2} respectively show the pdf $f(\tau | x_s)$ obtained through KS and the corresponding cdf for the same location and months of above, for $x_s=0$. Again, Fig.~\ref{fig:pdf_2} is consistent with the fact that in the summer days are longer and Fig.~\ref{fig:cdf_2} confirms the goodness of our KS estimation. Also in this case, the statistics for all months have passed the Kolmogorov-Smirnov test for a confidence of $1$\%. The pdfs for state $x_s=1$ are not shown as these are specular to those of Fig.~\ref{fig:pdf_2} and this is also to be expected as the sum of the duration of the two states $x_s=0$ (daytime) and $x_s=1$ (night) corresponds to the constant duration of a day. This means that the duration statistics of one state is sufficient to derive that of the other.\\

\noindent {\bf Slot-based clustering:} the attractive property of the $2$-state semi-Markov model obtained from the night-day clustering approach is its simplicity, as two states and four distributions suffice to statistically represent the energy inflow dynamics. Nevertheless, this model leads to a coarse-grained characterization of the temporal variation of the harvested current, especially in the high-energy state. 

\begin{figure}[t]
\begin{center}
\includegraphics[width=0.50\textwidth]{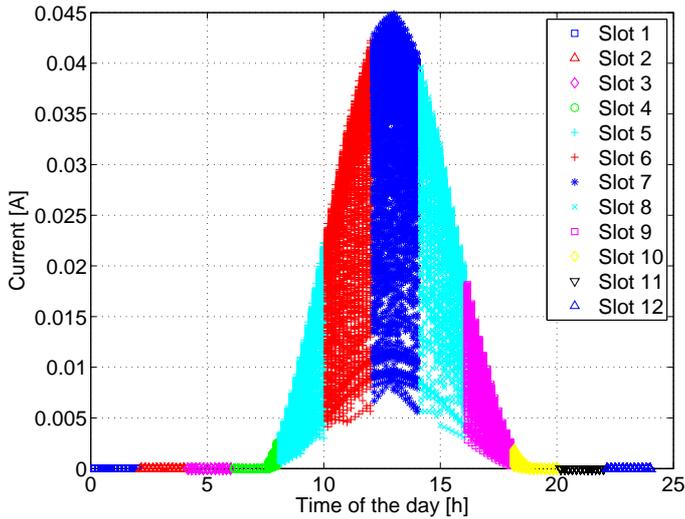}
\caption{Result of slot-based clustering considering $12$ time slots (states) for the month of July, years $1999-2010$.} 
\label{fig:slot-based-clustering-july}
\end{center}
\end{figure}

Slot-based clustering has been devised with the aim of capturing finer details. An example of the clustering result for this case is given in Fig.~\ref{fig:slot-based-clustering-july}, for the month of July. All slots in this case have the same duration, which has been fixed a priori and corresponds to $24/N_{\rm s}$~hours.

\begin{figure}[t]
 \begin{minipage}[b]{8.5cm}
   \centering
   \includegraphics[width=\textwidth]{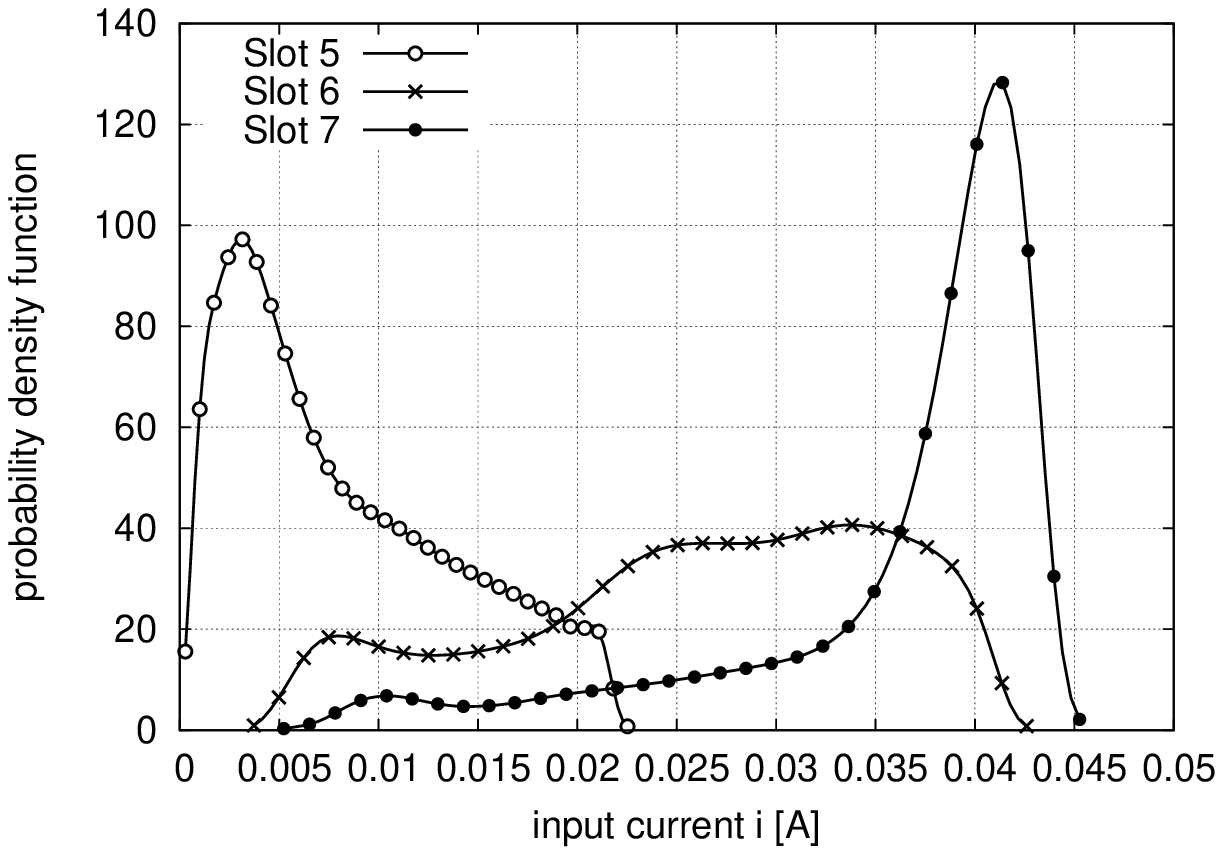}
   \caption{Probability density function (pdf) $g(i | x_s)$ for $x_s = 5,6$ and $7$ for the slot-based clustering method for the month of July.}
   \label{fig:pdf_slot}
 \end{minipage}
 \ \hspace{2mm} \hspace{3mm} \
 \begin{minipage}[b]{8.5cm}
  \centering
  \includegraphics[width=\textwidth]{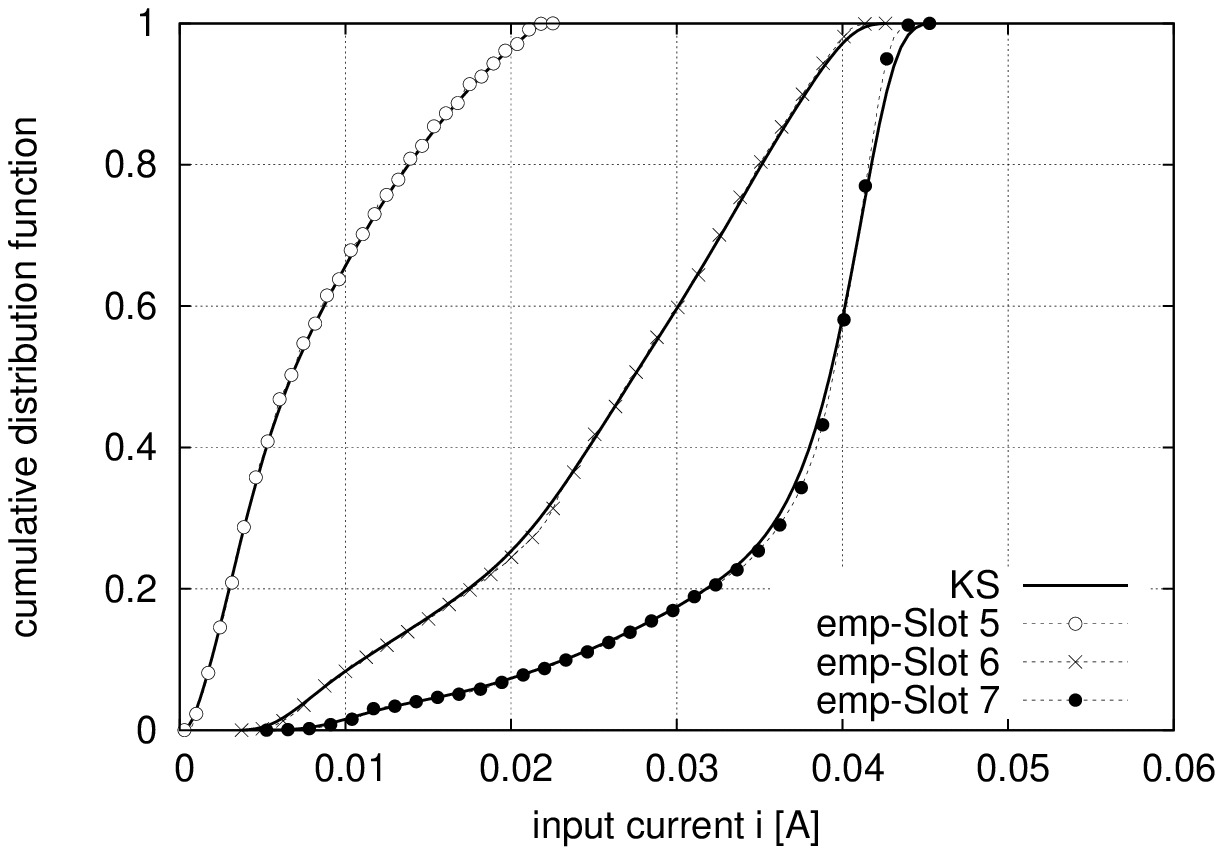}
   \caption{Comparison between KS and the empirical cdfs (emp) of the scavenged current for $x_s = 5,6$ and $7$ for the slot-based clustering method for the month of July.}
   \label{fig:cdf_slot}
 \end{minipage}
\end{figure}

Fig.~\ref{fig:pdf_slot} shows the pdf $g(i|x_s)$ for the first three states of the day (slots $5,6$ and $7$, see Fig.~\ref{fig:slot-based-clustering-july}) for the month of July, which have been obtained through KS. As expected, the peaks (and the supports) of the pdfs move to higher values, until reaching the maximum of $0.04$~A for slot $7$, which is around noon. Due to the symmetry in the solar distribution within the day, the results for the other daytime states are similar  and therefore have not been reported. In Fig.~\ref{fig:cdf_slot} we compare the cdfs obtained through KS against the empirical ones. Also in this case, all the cdfs have passed the Kolmogorov-Smirnov test for a confidence of $1$\%.

\begin{figure}[t]
\begin{center}
\includegraphics[width=\columnwidth]{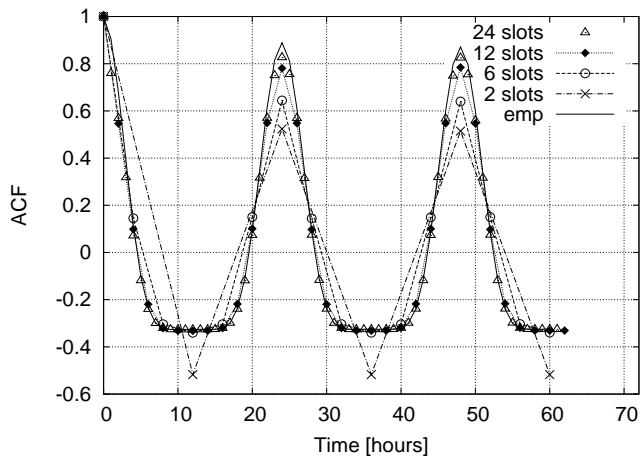}
\caption{Autocorrelation function for empirical data (solid curve) and for a synthetic Markov process generated through the night-day clustering ($2$ slots) and the slot-based clustering ($6$ and $12$ slots) approaches, obtained for the month of January.} 
\label{fig:acf}
\end{center}
\end{figure}

A last but important results is provided in Fig.~\ref{fig:acf}, where we plot the autocorrelation function (ACF) for the empirical data and the Markov processes obtained by slot-based clustering for a number of states $N_{\rm s}$ ranging from $2$ to $24$ for the month of January. With the ACF we test how well the Markov generated processes match the empirical data in terms of second-order statistics. As expected, a $2$-state Markov model poorly resembles the empirical ACF, whereas a Markov process with $N_{\rm s}=12$ states performs quite satisfactorily. Note also that $5$ of these $12$ states can be further grouped into a single macro-state, as basically no current is scavenged in any of them (see Fig.~\ref{fig:slot-based-clustering-july}). This leads to an equivalent Markov process with just eight states.

We highlight that our Markov approach keeps track of the temporal correlation of the harvested energy within the same day, though the Markovian energy generation process is independent of the ``day type'' (e.g., sunny, cloudy, rainy, etc.) and also on the previous day's type. Given this, one may expect a good fit of the ACF within a single day but a poor representation accuracy across multiple days. Instead, Fig.~\ref{fig:acf} reveals that the considered Markov modeling approach is sufficient to accurately represent second-order statistics. 
This has been observed for all months. 
On the other hand, one may be thinking of extending the state space by additionally tracking good ($g$) and bad ($b$) days so as to also model the temporal correlation associated with these qualities. This would amount to defining a Markov chain with the two macro-states $g$ and $b$, where $p_{gb} = \textrm{Prob}\{\textrm{day } k \textrm{ is } g | \textrm{ day } k-1 \textrm{ is } b \}$, with $k \geq 1$. Hence, in each state $g$ or $b$, the energy process could still be tracked according to one of the two clustering approaches of Section~\ref{sec:energy_model}, where the involved statistics would be now conditioned on being in the macro-state. The good approximation provided by our model, see Fig.~\ref{fig:acf}, show that this further level of sophistication is unnecessary.\\

\noindent {\bf Panel size and location:} to conclude, we show some illustrative results for different solar panel sizes and locations. Table~\ref{table:panelsize} presents the main outcomes for different solar cells configurations for the night-day clustering approach. Two representative months are considered: the month with the highest energy harvested, August, and the one with the lowest, December. As expected, the current inflow strongly depends on the panel size (linearly). Also, note that the day duration slightly increases for an increasing panel area as this value is obtained by measuring when the energy is above a certain (clustering) threshold. Although we scaled this threshold proportionally with an increasing harvested current, the longer duration of the day is due to the exponential behavior introduced by the scaling factor in \eq{eq:IVcurve}, see the RHS of this equation.

Finally, in table~\ref{table:panel_location} we show some energy harvesting figures for a solar panel with $n_{\rm p} = n_{\rm s} = 6$ for some representative cities.

\begin{table*}[tb]\footnotesize
  \caption{Results for different solar panel configurations with night-day clustering in Los Angeles}
\begin{center}
\begin{tabular}{c|c|c|c|c|c|c|c|c|c|c|c}
\toprule
 &  & \multicolumn{5}{c | }{Aug} & \multicolumn{5}{c}{Dec} \\
$n_{\rm p} \times n_{\rm s}$ & Size & $\bar{i}$ & $\max(i)$ & $\bar{\tau}$ & $\min(\tau)$ &  $\max(\tau)$ & $\bar{i}$ & $\max(i)$ & $\bar{\tau}$ & $\min(\tau)$ &  $\max(\tau)$\\
 & [cm$^2$] & [A] & [A] & [h] & [h] &  [h] & [A] & [A] & [h] & [h] & [h] \\
\midrule
2 x 2 & 2.99 & 0.002163 & 0.004524 & 9.73 & 8.17 & 10.17 & 0.001110 & 0.002484 & 7.74 & 5.00 & 8.33 \\
4 x 4 & 11.98 & 0.009254 & 0.019766 & 10.18 & 9.00 & 10.67 & 0.004847 & 0.011029 & 8.27 & 6.50 & 8.67 \\
6 x 6 & 26.96 & 0.021292 & 0.045561 & 10.26 & 9.17 & 10.67 & 0.011189 & 0.025666 & 8.38 & 6.67 & 8.83 \\
8 x 8 & 47.92 & 0.038149 & 0.082101 & 10.32 & 9.17 & 10.83 & 0.020115 & 0.046006 & 8.42 & 6.83 & 8.83 \\
10 x 10 & 74.88 & 0.059967 & 0.129194 & 10.34 & 9.17 & 10.83 & 0.031650 & 0.072437 & 8.44 & 6.83 & 8.83 \\
12 x 12 & 107.83 & 0.086729 & 0.186905 & 10.35 & 9.17 & 10.83 & 0.045795 & 0.104829 & 8.45 & 6.83 & 9.00 \\
\bottomrule
\end{tabular}
\end{center}
\label{table:panelsize}
\end{table*}

\begin{table*}[tb]\footnotesize
  \caption{Results for different solar panel locations for $n_{\rm p}=n_{\rm s}=6$}
\begin{center}
\begin{tabular}{l|c|c|c|c|c|c|c|c|c|c}
\toprule
  & \multicolumn{5}{c | }{Aug} & \multicolumn{5}{c}{Dec} \\
Location & $\bar{i}$ & $\max(i)$ & $\bar{\tau}$ & $\min(\tau)$ &  $\max(\tau)$ & $\bar{i}$ & $\max(i)$ & $\bar{\tau}$ & $\min(\tau)$ &  $\max(\tau)$\\
 & [A] & [A] & [h] & [h] &  [h] & [A] & [A] & [h] & [h] & [h] \\
\midrule
Chicago, IL & 0.017029 & 0.046742 & 10.57 & 8.50 & 11.33 & 0.005241 & 0.016084 & 6.95 & 4.83 & 8.00 \\
Los Angeles, CA & 0.021292 & 0.045561 & 10.26 & 9.17 & 10.67 & 0.011189 & 0.025666 & 8.38 & 6.67 & 8.83 \\
New York, NY & 0.017174 & 0.044617 & 10.42 & 8.83 & 11.00 & 0.006813 & 0.018945 & 7.57 & 5.67 & 8.33 \\
Reno, NV & 0.022912 & 0.048525 & 10.72 & 9.16 & 11.00 & 0.008247 & 0.021128 & 7.85 & 6.00 & 8.50 \\
\bottomrule
\end{tabular}
\end{center}
\label{table:panel_location}
\end{table*}

\section{Concluding Remarks}
\label{sec:conclusions}

In this paper we have considered micro-solar power sources, providing a methodology to model the energy inflow as a function of time through stochastic Markov processes. The latter, find application in energy self-sustainable systems, such as, for instance, in the simulation of energy harvesting communication networks and are as well useful to extend current theoretical work through more realistic energy models. Our approach has been validated against real energy traces, showing good accuracy in their statistical description in terms of first and second order statistics. 

Our tool has been developed using Matlab\texttrademark  and is available under the GPL license at~\cite{Solar_stat_code}.

\section*{Acknowledgment}

The research leading to the results in this paper has received funding from the Seventh Framework Programme (FP7/2007-2013) under grant agreement no. 251557 (Project SWAP) and no. 317762 (Project COMBO), by the Spanish Ministry of Science and Innovation under grant TEC2011-29700-C02-01 (Project SYMBIOSIS) and TEC2010-20823 (Project CO2GREEN) and by the Generalitat de Catalunya under grant 2009-SGR-940.

\bibliographystyle{IEEEtran}

\bibliography{EnergyCon_2014_SolarStat}

\enlargethispage{-40mm}

\end{document}